\documentclass[12pt]{article}

\usepackage[english]{babel}
\usepackage[cp1251]{inputenc}
\usepackage[T1]{fontenc}

\pdfoutput=1
\usepackage{makeidx}
\usepackage{amssymb}
\usepackage{amsfonts}
\usepackage{amsmath}
\usepackage{graphicx}
\usepackage{setspace}
\usepackage{authblk}
\usepackage{units}
\usepackage{appendix}
\usepackage{xparse}
\usepackage{cite}
\usepackage{hyperref}
\usepackage[left=2.1cm,top=2.5cm,right=2.1cm]{geometry}

\begin{document}

\title{Resonant entanglement of photon beams by a magnetic field}
\author{A. I. Breev$^{1}$\thanks{
		breev@mail.tsu.ru}, D. M. Gitman$^{1,2,3}$\thanks{
		dmitrygitman@hotmail.com}, \\
	$^{1}$\small{Department of Physics, Tomsk State University, \\
		Lenin ave. 36, 634050 Tomsk, Russia;}\\
	$^{2}\ $P.N. Lebedev Physical Institute, \\
	53 Leninskiy ave., 119991 Moscow, Russia.\\
	$^{3}$ Institute of Physics, University of S\~{a}o Paulo, \\
	Rua do Mat\~{a}o, 1371, CEP 05508-090, S\~{a}o Paulo, SP, Brazil.}

\maketitle

\abstract{
	In spite of the fact that photons do not interact with an external magnetic
	field, the latter field may indirectly affect photons in the presence of a
	charged environment. This opens up an interesting possibility to
	continuously control the entanglement of photon beams without using any
	crystalline devices. We study \ this possibility in the framework of an
	adequate QED model. In an approximation it was discovered that such
	entanglement has a resonant nature, namely, a peak behavior at certain
	magnetic field strengths, depending on characteristics of photon beams
	direction of the magnetic field and parameters of the charged medium.
	Numerical calculations illustrating the above-mentioned resonant behavior of
	the entanglement measure and some concluding remarks are presented.
	
}



\section{Introduction\label{Intr}}

Entanglement phenomenon is associated with a quantum non-separability of
parts of a composite system. Entangled states appear in studying principal
questions of quantum theory, they are considered as key elements in quantum
information theory in quantum computations and quantum cryptography
technologies; see e.g. Refs. \cite{Bell,NieCh00}. In laboratory conditions
the entanglement of photon beams is usually created and studied using some
kind of crystalline devices. In spite of the fact that photons do not
interact with an external magnetic field, the latter field may indirectly
affect photons in the presence of a charged environment. This opens up an
interesting possibility to continuously control the entanglement of photon
beams. Studying this possibility in the framework of an adequate QED model,
we have discovered that such entanglement has a resonant nature, namely, a
peak behavior at certain magnetic field strengths depending on
characteristics of photon beams and parameters of the charged medium. This
is the study presented in this article. The article is organized as follows:
In Sec. \ref{S1}, we outline details of the above-mentioned QED model. This
model describes a photon beam that consists of photons with two different
frequencies, moving in the same direction and interacting with a quantized
charged scalar field (KG field) placed in a constant magnetic field.
Particles of the KG field we call electrons in what follows and the totality
of the electrons is called the electron medium. Photons with each frequency
may have two possible linear polarizations. In the beginning, we consider
the electron subsystem consisting of only one charged particle. Both
quantized fields (electromagnetic and the KG one) are placed in a box of the
volume $V=L^{3}$ and periodic conditions are supposed. We believe that in
this case the model already describes the photons interacting with many
identical electrons, and the quantity $\rho =V^{-1}$ may be interpreted as
the density of the electron medium. In this article, we essentially correct
exact solutions used in our previous consideration of similar models; see
Ref. \cite{360} and references there. In a certain approximation, solutions
of the model correspond to two independent subsystems, one of which is a
quasi-electron medium and another one is a set of some quasi-photons. In the
new solutions the orders of smallness of contributions to quasi-photon
states used in calculating the entanglement measures are accurately
determined and an adequate expression for the spectrum of quasi-electrons
derived. Namely, the latter made it possible to detect the resonant behavior
of the entanglement measure at some resonant values of the external magnetic
field. Finally, in Sec. \ref{S3}, numerical calculations illustrating the
above-mentioned resonant behavior of the entanglement measure and some
concluding remarks are presented. Technical details related to Hamiltonian
diagonalization are placed in the Appendix \ref{A}.

\section{QED model and its solutions\label{S1}}

Consider photons with two different momenta $\mathbf{k}_{s}=\kappa _{s}
\mathbf{n}$, $s=1,2$ (frequencies), moving in the same direction $\mathbf{n}
=\left( 0,0,1\right)$ and interacting with quantized charged scalar
particles-electrons placed in a constant magnetic field $\mathbf{B}=B\mathbf{n}$,
$B>0$, potentials of which in the Landau gauge are: $\mathbf{A}_{\mathrm{ext}}(
\mathbf{r})=\left(-Bx^{2},0,0\right)$. In what follows, we use the system
of units where $\hbar=c=1$. The operator potentials $\hat{A}^{\mu }\left(
\mathbf{r}\right) $, $\mu =0,\dots,3$; $\mathbf{r}=\left(
x^{1},x^{2},x^{3}=z\right)$ of the photon beam are chosen in the Coulomb
gauge, $\hat{A}^{\mu}(\mathbf{r}) =( 0,\mathbf{\hat{A}}
(\mathbf{r}))$, $\mathrm{div}\mathbf{\hat{A}}(\mathbf{r})=0$,
in fact, they depend only on $z$,
\begin{eqnarray}
	&&\mathbf{\hat{A}}\left( \mathbf{r}\right) =\sum_{s=1,2}\sum_{\lambda =1,2}
	\sqrt{\frac{1}{2\kappa _{s}V}}\left[ \hat{a}_{s\mathbf{,}\lambda }\exp
	\left( i\kappa _{s}z\right) +\hat{a}_{s\mathbf{,}\lambda }^{\dagger }\exp
	\left( -i\kappa _{s}z\right) \right] \mathbf{e}_{\lambda }\ ,  \notag \\
	&&\left[ \hat{a}_{s\mathbf{,}\lambda },\hat{a}_{s^{\prime },\lambda ^{\prime
	}}\right] =0,\ \left[ \hat{a}_{s\mathbf{,}\lambda },\hat{a}_{s^{\prime
		},\lambda ^{\prime }}^{\dagger }\right] =\delta _{s,s^{\prime }}\delta
	_{\lambda ,\lambda ^{\prime }},\ \ s,s^{\prime }=1,2;\ \ \lambda ,\lambda
	^{\prime }=1,2\ .
	\label{1}
\end{eqnarray}
Here $\hat{a}_{s,\lambda }\ $and $\hat{a}_{s,\lambda }^{\dagger }$ are
creation and annihilation operators of the free photons from the beam, 
$\mathbf{e}_{\lambda}$ are real polarization vectors, $(\mathbf{e}_{\lambda}
\mathbf{e}_{\lambda ^{\prime }})=\delta _{\lambda ,\lambda ^{\prime }}$, $(
\mathbf{ne}_{\lambda })=0$. We choose the polarization vector in the form $
\mathbf{e}_{1}=(1,0,0)$, $\mathbf{e}_{2}=(0,1,0)$. The photon Fock space $
\mathfrak{H}_{\mathrm{\gamma }}$ is constructed by the creation and
annihilation operators and by the vacuum vector $\left\vert 0\right\rangle _{
\mathrm{\gamma }}$, $\hat{a}_{s,\lambda }\left\vert 0\right\rangle _{\mathrm{
\gamma }}=0$, $\forall s,\lambda$. Photon vectors are denoted as $
\left\vert \Psi \right\rangle _{\mathrm{\gamma }}$,$\ \left\vert \Psi
\right\rangle _{\mathrm{\gamma }}\in \mathfrak{H}_{\mathrm{\gamma }}$. The
Hamiltonian of free photon beam reads:
\begin{equation}
	\hat{H}_{\mathrm{\gamma }}=\sum_{s=1,2}\sum_{\lambda =1,2}\kappa _{s}\hat{a}
	_{s,\lambda }^{\dagger }\hat{a}_{s,\lambda },\ \kappa _{s}=\kappa
	_{0}d_{s},\ \kappa _{0}=2\pi L^{-1},\ d_{s}\in \mathbb{N},\ s=1,2\ .
	\label{2}
\end{equation}

Electrons are described by a scalar field $\varphi \left( \mathbf{r}\right) $
interacting with the external constant magnetic field $A_{\mathrm{ext}}^{\mu
}(\mathbf{r})$. The magnetic field does not violate the vacuum stability.
After the canonical quantization, the scalar field and its canonical
momentum $\pi (\mathbf{r})$ become operators $\hat{\varphi}(\mathbf{r})$ and 
$\hat{\pi}(\mathbf{r}).$ The corresponding Heisenberg operators $\hat{\varphi
}(x)$ and $\hat{\pi}(x)$, $x=(x^{\mu })=(t,\mathbf{r})$, satisfy the
equal-time nonzero commutation relations $[\hat{\varphi}(x),\hat{\pi}
(x^{\prime })]_{t=t^{\prime }}=i\delta (\mathbf{r}-\mathbf{r}^{\prime })$.
These operators act in the electron Fock space $\mathfrak{H}_{\mathrm{e}}$
constructed by a set of creation and annihilation operators of the scalar
particles and by a corresponding vacuum vector $\left\vert 0\right\rangle _{
\mathrm{e}}$. Electron vectors are denoted as $\left\vert \Psi \right\rangle
_{\mathrm{e}}$, $\left\vert \Psi \right\rangle _{\mathrm{e}}\subset 
\mathfrak{H}_{\mathrm{e}}$.

The Fock space $\mathfrak{H}$ of the complete system is a tensor product of
the photon Fock space and the electron Fock space, $\mathfrak{H=H}_{\mathrm{
\gamma }}\otimes \mathfrak{H}_{\mathrm{e}}$. Vectors from the Fock space 
$\mathfrak{H}$ are denoted by $\left\vert \Psi \right\rangle $, $\left\vert
\Psi \right\rangle \in \mathfrak{H}$.

The Hamiltonian of the complete system (composed of the photon and the
electron subsystems) has the following form:
\begin{align}
	& \hat{H}=\int \left\{ \hat{\pi}^{+}(\mathbf{r})\hat{\pi}(\mathbf{r})+\hat{
		\varphi}^{+}\left( \mathbf{r}\right) \left[ \mathbf{\hat{P}}^{2}(\mathbf{r}
	)+m^{2}\right] \hat{\varphi}\left( \mathbf{r}\right) \right\} d\mathbf{r}+
	\hat{H}_{\mathrm{\gamma }}\ ,  \notag \\
	& \mathbf{\hat{P}}(\mathbf{r})=\mathbf{\hat{p}}+e\left[ \mathbf{\hat{A}(}z)+
	\mathbf{A_{\mathrm{ext}}(r})\right] ,\ \mathbf{\hat{p}}=-i\nabla,\ e>0\ .  
	\label{3}
\end{align}%
Consider the amplitude-vector (AV) $\varphi (x)=\ _{\mathrm{e}}\left\langle
0\right\vert \hat{\varphi}(\mathbf{r})\left\vert \Psi \left( t\right)
\right\rangle$, which is on the one side a function on $x$ (the projection
of a vector $\left\vert \Psi \left( t\right) \right\rangle$ onto a
one-electron state), on the other side AV is a vector in the photon Fock
space. In the similar manner, one could introduce many-electron or positron
amplitudes and interpreted them as AVs of photons interacting with many
charged particles. However, we neglect the existence of such amplitudes in
the accepted further approximation, they are related to processes of virtual
pair creation. In such an approximation, one can demonstrate that AV 
$\varphi (x)$ satisfies the following equation:
\begin{equation}
	\left[ (i\partial _{t}-\hat{H}_{\mathrm{\gamma }})^{2}-\mathbf{\hat{P}}^{2}(
	\mathbf{r})-m^{2}\right] \varphi (x)=0\ .
	\label{4}
\end{equation}

It is convenient to pass from the AV $\varphi (x)$ to a AV $\Phi (x)=U_{
\mathrm{\gamma }}\left( t\right) \varphi (x)$, $U_{\mathrm{\gamma }}\left(
t\right) =\exp( i\hat{H}_{\mathrm{\gamma }}t)$, which satisfies
a KG like equation (KGE):
\begin{align}
	& \left[ \hat{P}_{\mu }\hat{P}^{\mu }-m^{2}\right] \Phi (x)=0,\ \hat{P}^{\mu
	}=\hat{P}^{\mu }\left( x\right) =i\partial ^{\mu }+e\left[ \hat{A}^{\mu
	}(u)+A_{\mathrm{ext}}^{\mu }(\mathbf{r})\right] \ ,  \notag \\
	& \hat{A}^{\mu }(u)=\left( 0,\mathbf{\hat{A}}(u)\right) ,\ u=t-z,\ \mathbf{%
		\hat{A}}(u)=U_{\mathrm{\gamma }}\left( t\right) \mathbf{\hat{A}}(z)U_{%
		\mathrm{\gamma }}^{-1}\left( t\right)  \notag \\
	& \ =\frac{1}{e}\sum_{s=1,2}\sum_{\lambda =1,2}\sqrt{\frac{\varepsilon }{%
			2\kappa _{s}}}\left[ \hat{a}_{s,\lambda }\exp \left( -i\kappa _{s}u\right) +%
	\hat{a}_{s,\lambda }^{\dagger }\exp \left( i\kappa _{s}u\right) \right] 
	\mathbf{e}_{\lambda }\ ,  \label{5}
\end{align}%
where $\varepsilon =\alpha \rho $,$\ \alpha =e^{2}/\hbar c=1/137$, and $\rho 
$ is the density of the electron media. The quantity $\varepsilon $
characterizes the strength of the interaction between the charged particles
and the photon beam. We suppose that both $\varepsilon $ and $\alpha $ are
small, this supposition defines the above mentioned approximation.

One can see that in the model under consideration, we have three commuting
integrals of motion $\hat{G}_{\mu }=i\partial _{\nu }+n_{\nu }\hat{H}_{
\mathrm{\gamma }}$, $\mu =0$,$1$,$3$; $n^{\mu }=(1,\mathbf{n})$; $\hat{G}
_{0} $ can be interpreted as the operator of the total energy and $\hat{G}
_{\mu}$, $\mu =1$, $3$ as momenta operators in the directions $x^{1}$ and 
$z$.

Recall that $\hat{I}$ is an integral of motion if its mean value
\begin{equation}
	\left( \varphi ,\hat{I}\varphi \right) =\int \varphi ^{\ast }\left( x\right)
	\left( i\overleftrightarrow{\partial _{0}}-2eA_{0}\right) \hat{I}\varphi
	\left( x\right) d\mathbf{r\,},\ \overleftrightarrow{\partial _{0}}=
	\overrightarrow{\partial _{0}}-\overleftarrow{\partial _{0}}\ ,  \notag
\end{equation}%
with respect to any $\varphi$ satisfying the KGE does not depend on time.
If $\hat{I}$ is an integral of motion, then $\left[ \hat{I},\hat{P}_{\mu}
\hat{P}^{\mu }\right]=0$.

If $\hat{I}$ is an integral of motion, then, apart from satisfying the KGE,
the wave function could be choose as an eigenfunction of $\hat{I}$. Then we
look for AV $\Phi (x)$ that are also eigenvectors for the integrals of
motion $\hat{G}_{\mu}$,
\begin{eqnarray}
	&&\left[ \hat{P}_{\mu }\hat{P}^{\mu }-m^{2}\right] \Phi (x)=0\ ,  \notag \\
	&&\hat{G}_{\mu }\Phi (x)=g_{\mu }\Phi (x),\ \ \mu =0,1,3\ ,
	\label{6b}
\end{eqnarray}%
where $g_{0}$ is the total energy and $g_{1,3}$ are momenta in $x^{1}$ and 
$z$ directions. From Eq. (\ref{6b}) it follows
\begin{eqnarray}
	&&\left[ \hat{P}_{\mu }\hat{P}^{\mu }-m^{2}\right] \Phi (x)=\frac{1}{2(ng)}[%
	\hat{H}_{\mathrm{\chi }}\left( u\right) -(g_{0}-g_{3})/2]\Phi (x),\ \
	(ng)=g_{0}+g_{3},  \notag \\
	&&\hat{H}_{\mathrm{\chi }}\left( u\right) =\hat{H}_{\mathrm{\gamma }}+\frac{1%
	}{2(ng)}\left\{ \left[ eBx^{2}-g^{1}-e\hat{A}^{1}\left( u\right) \right]
	^{2}+\left[ i\partial _{2}-eA^{2}\left( u\right) \right] ^{2}+m^{2}\right\}
	\ .  
	\label{d.7}
\end{eqnarray}
Consequently, the operator $\hat{H}_{\mathrm{\chi }}\left( u\right)$ 
commutes with the operator $\hat{P}_{\mu }\hat{P}^{\mu }$ on solutions 
$\Phi(x)$, and therefore is an integral of motion.

A solution to Eq. (\ref{6b}) has form
\begin{equation}
	\Phi (x)=\exp \left[ -i(g_{0}t+g_{1}x^{1}+g_{3}z)\right] \hat{U}_{\mathrm{
	\gamma }}\left( u\right) \chi (x^{2})\ ,  
	\label{d.8}
\end{equation}
where the function $\chi (x^{2})$ must satisfy the following equation:
\begin{equation}
	\hat{H}_{\mathrm{\chi }}(0)\chi (x^{2})=\frac{g_{0}-g_{3}}{2}\chi (x^{2})\ .
	\label{8}
\end{equation}

In order to solve the latter equation, we pass to a description of the
electron motion in the magnetic field in an adequate Fock space, 
see Ref. \cite{154}. We introduce new creation $\hat{a}_{0}^{\dagger }$ and
annihilation $\hat{a}_{0}$ Bose operators, $[\hat{a}_{0},\hat{a}
_{0}^{\dagger }]=1$,
\begin{align}
	& \hat{a}_{0}=(2)^{-1/2}\left( \eta +\partial _{\eta }\right) ,\ \ \hat{a}
	_{0}^{\dagger }=(2)^{-1/2}\left( \eta -\partial _{\eta }\right) ,\
	(eB)^{1/2}\eta =\left( eBx^{2}-g^{1}\right) \ ,  \notag \\
	& x^{2}=\sqrt{\frac{1}{2eB}}\left( \hat{a}_{0}-\hat{a}_{0}^{\dagger }\right)
	+g^{1},\ \ \partial _{2}=\sqrt{\frac{eB}{2}}\left( \hat{a}_{0}+\hat{a}
	_{0}^{\dagger }\right) \ ,\ \ \hat{a}_{0}\left\vert 0\right\rangle _{\mathrm{
			e}}=0\ .  
	\label{a1}
\end{align}
These operators commute with all the photon operators $a_{s,\lambda
}^{\dagger }$ and $\hat{a}_{s,\lambda }$, $s=1,2$, $\lambda =1,2$. We denote
the totality of the free photon and the introduced electron creation and
annihilation operators as $a_{s,\lambda }^{\dagger }$ and $\hat{a}
_{s,\lambda }$, $s=0,1,2$, where $\hat{a}_{0,\lambda }^{\dagger }=\hat{a}
_{0}^{\dagger }\delta _{\lambda ,1}$ and $\hat{a}_{0,\lambda }=\hat{a}
_{0}\delta _{\lambda ,1}$. The corresponding vacuum vector $\left\vert
0\right\rangle$ reads: 
\begin{equation}
	\left\vert 0\right\rangle =\left\vert 0\right\rangle _{\mathrm{\gamma }
	}\otimes \left\vert 0\right\rangle _{\mathrm{e}},\ \hat{a}_{s,\lambda
	}\left\vert 0\right\rangle =0,\ \ s=0,1,2\ .  
	\label{a1a}
\end{equation}

The operator $\hat{H}_{\mathrm{\chi }}\left( 0\right) $ can be represented
as a quadratic form in terms this totality of the creation and annihilation
operators,
\begin{eqnarray}
	&&\hat{H}_{\mathrm{\chi }}(0)=\sum_{s,s^{\prime }=0,1,2}\sum_{\lambda
		,\lambda ^{\prime }=1,2}A_{s,\lambda ;s^{\prime },\lambda ^{\prime
	}}a_{s,\lambda }^{\dagger }\hat{a}_{s^{\prime },\lambda ^{\prime }}+\frac{1}{%
		2}\left( B_{s,\lambda ;s^{\prime },\lambda ^{\prime }}\hat{a}_{s,\lambda
	}^{\dagger }\hat{a}_{s^{\prime },\lambda ^{\prime }}^{\dagger }+B_{s,\lambda
		;s^{\prime },\lambda ^{\prime }}^{\ast }\hat{a}_{s,\lambda }\hat{a}%
	_{s^{\prime },\lambda ^{\prime }}\right)  \notag \\
	&&+\frac{m^{2}}{2ng}+\frac{\omega }{2}+\frac{\epsilon }{2}\sum_{s=1,2}\kappa
	_{s}^{-1}\ ,\ \omega =\frac{eB}{ng},\ \epsilon =\frac{\varepsilon }{ng}\ , 
	\notag \\
	&&A_{s,\lambda ;s^{\prime },\lambda ^{\prime }}=\left[ \omega (2-\lambda
	)\delta _{0,s}+\kappa _{s}\left( 1-\delta _{0,s}\right) \right] \delta
	_{s,s^{\prime }}\delta _{\lambda ,\lambda ^{\prime }}+\frac{\epsilon }{2%
		\sqrt{\kappa _{s}\kappa _{s^{\prime }}}}\left( 1-\delta _{0,s}\right) \left(
	1-\delta _{0,s^{\prime }}\right) \delta _{\lambda ,\lambda ^{\prime }} 
	\notag \\
	&&-\frac{\left\vert ng\right\vert }{ng}\frac{\sqrt{\epsilon \omega }}{2}%
	\left[ \frac{(+i)^{\lambda -1}}{\sqrt{\kappa _{s}}}\left( 1-\delta
	_{0,s}\right) \delta _{0,s^{\prime }}\delta _{\lambda ^{\prime },1}+\frac{%
		(-i)^{\lambda ^{\prime }-1}}{\sqrt{\kappa _{s^{\prime }}}}\left( 1-\delta
	_{0,s^{\prime }}\right) \delta _{0,s}\delta _{\lambda ,1}\right] \ ,  \notag
	\\
	&&B_{s,\lambda ;s^{\prime },\lambda ^{\prime }}=\frac{\epsilon }{2\sqrt{%
			\kappa _{s}\kappa _{s^{\prime }}}}\left( 1-\delta _{0,s}\right) \left(
	1-\delta _{0,s^{\prime }}\right) \delta _{\lambda ,\lambda ^{\prime }} 
	\notag \\
	&&-\frac{\left\vert ng\right\vert }{ng}\frac{\sqrt{\epsilon \omega }}{2}%
	\left[ \frac{(-i)^{\lambda -1}}{\sqrt{\kappa _{s}}}\left( 1-\delta
	_{0,s}\right) \delta _{0,s^{\prime }}\delta _{\lambda ^{\prime },1}+\frac{%
		(-i)^{\lambda ^{\prime }-1}}{\sqrt{\kappa _{s^{\prime }}}}\left( 1-\delta
	_{0,s^{\prime }}\right) \delta _{0,s}\delta _{\lambda ,1}\right] \ .
	\label{a2}
\end{eqnarray}

As it is demonstrated in Appendix \ref{A}, there exists a linear canonical
transformation of the operators $a_{s,\lambda }^{\dagger }$ and $\hat{a}
_{s,\lambda }$, $s=0,1,2$, given by Eqs. (\ref{ac1}) which diagonalizes the
Hamiltonian $\hat{H}_{\mathrm{\chi }}\left( 0\right)$,
\begin{align}
	& \hat{H}_{\mathrm{\chi }}\left( 0\right) =\hat{H}_{\mathrm{e}}(0)+\hat{H}_{
		\mathrm{q-ph}}(0)\ \mathrm{,\ }\hat{H}_{\mathrm{e}}(0)=\tau
	_{0}c_{0}^{\dagger }c_{0}+\frac{m^{2}}{2(ng)}+\frac{\omega }{2}\ ,  \notag \\
	& \hat{H}_{\mathrm{q-ph}}(0)=\sum_{s=1,2}\sum_{\lambda =1,2}\tau _{s,\lambda
	}c_{s,\lambda }^{\dagger }c_{s,\lambda }\ -\sum_{s,k=0,1,2}\sum_{\lambda
		,\lambda ^{\prime }=1,2}\tau _{k,\lambda ^{\prime }}\left\vert v_{s,\lambda
		;k,\lambda ^{\prime }}\right\vert ^{2}+\frac{\epsilon }{2}\sum_{s=1,2}\kappa
	_{s}^{-1}\ ,  
	\label{13a}
\end{align}
where the quantities $\tau _{k,\lambda }$ satisfy the conditions $\tau
_{0}(\epsilon =0)=\omega $, $\tau _{k,\lambda }(\epsilon =0)=\kappa_{k}$
being positive roots of the equation
\begin{equation}
	\sum_{s=1,2}\frac{\epsilon }{\tau _{k,\lambda }^{2}-\kappa _{s}^{2}}=1+\frac{
		\left( -1\right) ^{\lambda }\omega }{\tau _{k,\lambda }},\ \ \tau
	_{0,\lambda }=\tau _{0}\delta _{\lambda ,1},\ \ k=0,1,2\ .  
	\label{14}
\end{equation}

It is possible to demonstrate that after an unitary transformation, the
integral of motion $\hat{H}_{\mathrm{\chi }}\left( u\right)$ can be
separated in two parts $\hat{H}_{\mathrm{q-ph}}(u)$ and 
$\hat{H}_{\mathrm{e}}(u)$:
\begin{eqnarray}
	&&\hat{H}_{\mathrm{\chi }}(u)=\hat{U}_{\mathrm{\gamma }}\left( u\right) \hat{
		H}_{\mathrm{\chi }}\left( 0\right) \hat{U}_{\mathrm{\gamma }}\left( u\right)
	^{-1}=\hat{H}_{\mathrm{q-ph}}(u)+\hat{H}_{\mathrm{e}}(u)\ ,  \notag \\
	&&\hat{H}_{\mathrm{q-ph}}(u)=\hat{U}_{\mathrm{\gamma }}\left( u\right) \hat{H
	}_{\mathrm{q-ph}}(0)\hat{U}_{\mathrm{\gamma }}\left( u\right) ^{-1},\ \ \hat{
		H}_{\mathrm{e}}(u)=\hat{U}_{\mathrm{\gamma }}\left( u\right) \hat{H}_{
		\mathrm{e}}(0)\hat{U}_{\mathrm{\gamma }}\left( u\right) ^{-1}\ ,  \notag \\
	&&\left[ \hat{H}_{\mathrm{e}}(u),\hat{H}_{\mathrm{\chi }}(u)\right] =\left[ 
	\hat{H}_{\mathrm{q-ph}}(u),\hat{H}_{\mathrm{\chi }}(u)\right] =0\ .
	\label{12}
\end{eqnarray}
Each of these parts are also integrals of motion due to relations Eqs. 
(\ref{d.7}), (\ref{8}) and (\ref{12}). The operator $\hat{H}_{\mathrm{e}}$
corresponds to the quasi-electron subsystem, while the operator 
$\hat{H}_{\mathrm{q-ph}}$ to the subsystem of quasi-photons.

It is useful to consider operators $\mathcal{\hat{P}}_{\mu }$,
\begin{eqnarray}
	&&\mathcal{\hat{P}}_{\mu }=\hat{G}_{\mu }\ -n_{\mu }\hat{H}_{\mathrm{q-ph}
	}\left( u\right) =i\partial _{\mu }-n_{\mu }\left[ \hat{H}_{\mathrm{q-ph}
	}\left( u\right) -\hat{H}_{\mathrm{\gamma }}\right] ,  \notag \\
	&&\left[ \mathcal{\hat{P}}_{\mu },\mathcal{\hat{P}}_{\nu }\right] =0,\ \ \mu
	,\nu =0,1,2\ ,  
	\label{12b}
\end{eqnarray}
witch are also integrals of motion. If we assume that at $\epsilon
\rightarrow 0$ the photons do not interact with the electronic medium, then
in such a limit the operators $\mathcal{\hat{P}}_{\mu }$ are the
energy-momentum operators of a the free electrons $i\partial _{\mu }$, and
the operator $n_{\mu }\hat{H}_{\mathrm{q-ph}}\left( u\right) $ is the
energy-momentum operator of the free photons $\hat{H}_{\mathrm{\gamma }}$.
It is therefore appropriate to refer to $\mathcal{\hat{P}}_{\mu }$ as the
quasi-electron energy-momentum, and to 
$n_{\mu }\hat{H}_{\mathrm{q-ph}}\left( u\right) $ as the energy-momentum of the quasi-photons.

Then we can choose AV $\Phi (x)$ to be eigenvectors for the integrals of
motion $\hat{H}_{\mathrm{e}}\left( u\right) $, $\hat{H}_{\mathrm{q-ph}}(u)$
and $\mathcal{\hat{P}}_{\mu }$,
\begin{equation}
	\hat{H}_{\mathrm{e}}(u)\Phi (x)=E_{\mathrm{e}}\Phi (x),\ \hat{H}_{\mathrm{
			q-ph}}(u)\Phi (x)=E_{\mathrm{q-ph}}\Phi (x),\ \mathcal{\hat{P}}_{\mu }\ \Phi
	(x)=p_{\mu }\Phi (x)\ .  
	\label{13}
\end{equation}
Further, we interpret the eigenvalues $p_{\mu }$ as momenta of
quasi-electrons. It follows from Eq. (\ref{13}) that $\Phi (x)$ is an
eigenvector for the operator $\hat{H}_{\mathrm{\chi }}(u)$,
\begin{equation}
	\hat{H}_{\mathrm{\chi }}(u)\Phi (x)=E\Phi (x),\ E=E_{\mathrm{e}}+E_{\mathrm{
			q-ph}}\ .  
	\label{13b}
\end{equation}

Substituting (\ref{d.8}) into Eq. (\ref{13b}), we obtain an equation for the
function $\chi (x^{2})$, 
\begin{equation}
	\hat{H}_{\mathrm{\chi }}(0)\chi (x^{2})=E\chi (x^{2})\ ,  
	\label{8b}
\end{equation}
which has the following solutions:
\begin{align}
	& \ \chi (x^{2})=\left\vert \phi _{\mathrm{e}}\right\rangle \otimes
	\left\vert \Phi _{\mathrm{q-ph}}\right\rangle ,\ \ \left\vert \Phi _{\mathrm{
			q-ph}}\right\rangle =\prod_{\lambda =1,2}\frac{\left( \hat{c}_{1,\lambda
		}^{\dagger }\right) ^{N_{1,\lambda }}}{\sqrt{N_{1,\lambda }!}}\left\vert
	0\right\rangle _{\mathrm{c}_{1}} \prod_{\lambda ^{\prime }=1,2}\frac{\left( \hat{c}_{2,\lambda
			^{\prime }}^{\dagger }\right) ^{N_{2,\lambda ^{\prime }}}}{\sqrt{
			N_{2,\lambda ^{\prime }}!}}\left\vert 0\right\rangle _{\mathrm{c}_{2}} \, , \nonumber\\
	&\left\vert \phi _{\mathrm{e}}\right\rangle =\frac{\left( \hat{c}
		_{0}^{\dagger }\right) ^{N_{0}}}{\sqrt{N_{0}!}}\left\vert 0\right\rangle _{
		\mathrm{c}_{0}},\ \hat{c}_{0}\left\vert 0\right\rangle _{c_{0}}=0,\ \
	c_{s,\lambda }\left\vert 0\right\rangle _{\mathrm{c}_{s}}=0\ , \quad
	E_{\mathrm{e}}=\tau _{0}N_{0}+\frac{m^{2}}{2(ng)}+\frac{\omega }{2},\ \
	N_{0}\in \mathbb{N}\ ,  \notag \\
	& E_{\mathrm{q-ph}}=\sum_{s=1,2}\sum_{\lambda =1,2}\tau _{s,\lambda
	}N_{s,\lambda }\nonumber\\
	&\qquad\,\,\,\,\, -\sum_{s,k=0,1,2}\sum_{\lambda ,\lambda ^{\prime }=1,2}\tau
	_{k,\lambda ^{\prime }}\left\vert v_{s,\lambda ;k,\lambda ^{\prime
	}}\right\vert ^{2}+\frac{\epsilon }{2}\sum_{s=1,2}\kappa _{s}^{-1},\ \
	N_{s,\lambda }\in \mathbb{N}\ .  
	\label{16}
\end{align}

Equations (\ref{13}),(\ref{8}) and (\ref{8b}) are consistent if
\begin{equation*}
	g_{0}=p_{0}+E_{\mathrm{q-ph}},\ \ g_{1}=p_{1},\ \ g_{3}=p_{3}-E_{\mathrm{q-ph
	}},\ \ E=\frac{g_{0}-g_{3}}{2}\ ,
\end{equation*}
which implies:
\begin{equation}
	E_{\mathrm{e}}+E_{\mathrm{q-ph}}\ =\frac{g_{0}-g_{3}}{2}=\frac{p_{0}-p_{3}}{2
	}+E_{\mathrm{q-ph}},\ \ \left( ng\right) =\left( np\right) \ ,  
	\label{10b}
\end{equation}
and
\begin{equation}
	E_{\mathrm{e}}=\frac{p_{0}-p_{3}}{2}\ .  
	\label{11b}
\end{equation}
Taking into account Eq. (\ref{16}) from (\ref{11b}) we obtain the spectrum
of quasi-electrons in the constant magnetic field:
\begin{equation}
	p_{0}^{2}=2eB\left( \frac{\tau _{0}}{\omega }N_{0}+\frac{1}{2}\right)
	+p_{3}^{2}+m^{2},\ \ \ \omega =eB(np)^{-1}\ .  
	\label{12c}
\end{equation}
Since $\tau _{0}(\epsilon =0)=\omega$, the well-known spectrum of a
relativistic spinless particle in the constant magnetic field, follows from
Eq. (\ref{12c}), 
\begin{equation}
	\bar{p}_{0}^{2}=2eB\left( N_{0}+\frac{1}{2}\right) +p_{3}^{2}+m^{2}\ .
	\label{12d}
\end{equation}

For small $\varepsilon $ the roots $\tau _{k,\lambda }$ are:%
\begin{eqnarray}
	&&\tau _{k,\lambda }=\kappa _{k}+\frac{\omega _{0}}{2eB\left( \kappa
		_{k}+(-1)^{\lambda }\omega _{0}\right) }\varepsilon +O(\varepsilon ^{2}),\ \
	k,\lambda =1,2\ ,  \label{15} \\
	&&\tau _{0}=\omega _{0}\left\{ 1-\varepsilon \frac{\omega _{0}}{eB}\left( 
	\frac{\omega _{0}N_{0}}{\bar{p}_{0}}-1\right) \sum_{s^{\prime }=1,2}\left(
	\omega _{0}^{2}-\kappa _{s^{\prime }}^{2}\right) ^{-1}\right\}
	+O(\varepsilon ^{2}),\ \omega _{0}=\frac{eB}{\bar{p}_{0}+p_{3}}\ .
	\label{15b}
\end{eqnarray}
In this approximation, the spectrum of the quasi-electrons in the constant
magnetic field has form:
\begin{equation}
	p_{0}=\bar{p}_{0}+\epsilon \frac{\omega _{0}N_{0}}{\bar{p}_{0}}\left( \frac{
		\omega _{0}N_{0}}{\bar{p}_{0}}-1\right) \sum_{s^{\prime }=1,2}\left( \omega
	_{0}^{2}-\kappa _{s^{\prime }}^{2}\right) ^{-1}+O(\varepsilon ^{2})\ .
	\label{16b}
\end{equation}
Using Eqs. (\ref{15}) and (\ref{15b}) we obtain for small $\varepsilon $
expressions for matrices (\ref{a9}) defining the canonical transformation 
(\ref{ac1}):
\begin{eqnarray}
	&&u_{s,\lambda ;k,\sigma }=-\left\{ \delta _{\lambda ,1}\delta _{s,0}\delta
	_{\lambda ,\sigma }\delta _{s,k}+\frac{\mathrm{sgn}\left[ \omega
		_{0}+(-1)^{\lambda }\kappa _{s}\right] }{\sqrt{2}}\left[ \delta _{\lambda
		,1}+i(-1)^{\sigma }\delta _{\lambda ,2}\right] \left( 1-\delta _{s,0}\right)
	\delta _{s,k}\right.  \notag \\
	&&\qquad\qquad\,\,+\left. \frac{\omega _{0}\sqrt{\varepsilon }}{\sqrt{2eB\kappa _{k}}%
		\left\vert \omega _{0}-\kappa _{k}\right\vert }\delta _{s,0}\left( 1-\delta
	_{k,0}\right) \delta _{\sigma ,1}\right\} +O(\varepsilon )\ ,  \notag \\
	&&v_{s,\lambda ;s^{\prime },\lambda ^{\prime }}=\left\{ \frac{(-i)^{\lambda
			-1}\omega _{0}\sqrt{\varepsilon }}{\sqrt{2eB\kappa _{s}}\left( \omega
		_{0}+\kappa _{s}\right) }\left( 1-\delta _{s,0}\right) \delta _{s,k}-\delta
	_{s,0}\left( 1-\delta _{k,0}\right) \delta _{\lambda ,1}\delta _{\sigma ,2}%
	\frac{\omega _{0}\sqrt{\varepsilon }}{\sqrt{2eB\kappa _{k}}\left( \omega
		_{0}+\kappa _{k}\right) }\right.  \notag \\
	&&\qquad\quad\,\,\,\,+\left. \sqrt{\frac{\kappa _{k}}{\kappa _{s}}}\frac{\omega _{0}\left[
		\delta _{\lambda ,1}+i(-1)^{\sigma }\delta _{\lambda ,2}\right] \varepsilon 
	}{2eB\sqrt{2}\left( \kappa _{s}+\kappa _{k}\right) \left\vert \omega
		_{0}+\left( -1\right) ^{\sigma }\kappa _{k}\right\vert }\left( 1-\delta
	_{s,0}\right) \left( 1-\delta _{k,0}\right) \right\} +O(\varepsilon )\ .
	\label{16c}
\end{eqnarray}

Substituting $\chi (x^{2})$ given by Eq. (\ref{16}) in equation (\ref{d.8})
for $\Phi (x)$, for small $\varepsilon $ we obtain:
\begin{eqnarray}
	&&\Phi (x)=\left\vert \Phi _{\mathrm{e}}\right\rangle \otimes \left\vert
	\Phi _{\mathrm{q-ph}}\right\rangle +O(\varepsilon )\ ,  \notag \\
	&&\left\vert \Phi _{\mathrm{e}}\right\rangle =\exp \left\{ -i\left(
	p_{0}t+p_{1}x^{1}+p_{3}z\right) \right\} \frac{\left( \hat{c}_{0}^{\dagger
		}\right) ^{N_{0}}}{\sqrt{N_{0}!}}\left\vert 0\right\rangle _{\mathrm{c}
		_{0}}\ .  
	\label{17}
\end{eqnarray}

\section{Photon entanglement problem\label{S2}}

\subsection{General\label{SS2.1}}

We recall that a qubit is a two-level quantum-mechanical system with state
vectors (two columns) $\left\vert \psi \right\rangle =\left( \psi _{1},\psi
_{2}\right) ^{T}\in $ $\mathcal{H}=\mathbb{C}^{2}$, $\langle \psi ^{\prime
}\left\vert \psi \right\rangle =\psi _{1}^{\prime \ast }\psi _{1}+\psi
_{2}^{\prime \ast }\psi _{2}$. An orthogonal basis $\left\vert
a\right\rangle $, $a=0$, $1$ in $\mathcal{H}$ is: $\left\vert
0\right\rangle =\left( 1,0\right) ^{T}$, $\left\vert 1\right\rangle =\left(
0,1\right) ^{T}$, $\langle a\left\vert a^{\prime }\right\rangle =\delta
_{aa^{\prime }}$, $\sum_{a=0,1}\left\vert a\right\rangle \langle a|\ =I$, 
where $I=\mathrm{diag}\left( 1,1\right)$. E.g. two levels can be taken as
spin up and spin down of an electron; or two polarizations of a single
photon. A system, composed of two qubit subsystems $A$ and $B$ with the
Hilbert space $\mathcal{H}_{AB}=\mathcal{H}_{A}\otimes \mathcal{H}_{B}$
where $\mathcal{H}_{A/B}=\mathbb{C}^{2}$ is a four level system. If 
$\left\vert a\right\rangle _{A}$ and $\left\vert b\right\rangle _{B}$,$\ a$, 
$b=0$, $1$, are orthonormal bases in $\mathcal{H}_{A}$ and $\mathcal{H}_{B}$
respectively, then $\left\vert \alpha b\right\rangle =\left\vert
a\right\rangle \otimes \left\vert b\right\rangle $ is a complete and
orthonormalized basis in $\mathcal{H}_{AB}$, $\left\vert \alpha
b\right\rangle =\left\vert a\right\rangle \otimes \left\vert b\right\rangle
=\left( a_{1}b_{1},a_{1}b_{2},a_{2}b_{1},a_{2}b_{2}\right) ^{T}$. The
so-called computational basis $\left\vert \Theta \right\rangle _{s}$, 
$s=1,2,3,4$, reads:
\begin{align}
	& \left\vert \Theta \right\rangle _{1}=\left\vert 00\right\rangle =\left( 
	\begin{array}{cccc}
		1 & 0 & 0 & 0
	\end{array}%
	\right) ^{T},\ \left\vert \Theta \right\rangle _{2}=\left\vert
	01\right\rangle =\left( 
	\begin{array}{cccc}
		0 & 1 & 0 & 0
	\end{array}%
	\right) ^{T}\ ,  \notag \\
	& \left\vert \Theta \right\rangle _{3}=\left\vert 10\right\rangle =\left( 
	\begin{array}{cccc}
		0 & 0 & 1 & 0
	\end{array}%
	\right) ^{T},\ \left\vert \Theta \right\rangle _{4}=\left\vert
	11\right\rangle =\left( 
	\begin{array}{cccc}
		0 & 0 & 0 & 1
	\end{array}%
	\right) ^{T}\ .  
	\label{18}
\end{align}

A pure state $\left\vert \Psi \right\rangle _{AB}\in \mathcal{H}_{AB}$ is
called \textbf{separable} iff it can be represented as: $\left\vert \Psi
\right\rangle _{AB}=\left\vert \Psi \right\rangle _{A}\otimes \left\vert
\Psi \right\rangle _{B}$,\ $\left\vert \Psi \right\rangle _{A}\in \mathcal{H}
_{A}$, $\left\vert \Psi \right\rangle _{B}\in \mathcal{H}_{B}$. Otherwise,
it is \textbf{entangled}. An entanglement measure $M\left( \left\vert \Psi
\right\rangle _{AB}\right)$ of the state $\left\vert \Psi \right\rangle
_{AB}$ is real and positive. This measure is zero for separable states, and
is $1$ for maximally entangled states. In what follows, we use the
information measure
\begin{align}
	& M\left( \left\vert \Psi \right\rangle _{AB}\right) =S\left( \hat{\rho}
	_{A}\right) =S\left( \hat{\rho}_{B}\right) \ ,\ S\left( \hat{\rho}
	_{A/B}\right) =-\mathrm{tr}\left( \hat{\rho}_{A/B}\log \hat{\rho}
	_{A/B}\right) \ ,  \notag \\
	& \hat{\rho}_{A}=\mathrm{tr}_{B}\hat{\rho}_{AB}=\sum_{b}\langle b\left\vert 
	\hat{\rho}_{AB}|b\right\rangle ,\ \ \hat{\rho}_{B}=\mathrm{tr}_{A}\hat{\rho}
	_{AB}=\sum_{a}\langle a\left\vert \hat{\rho}_{AB}|a\right\rangle \ ,
	\label{19}
\end{align}
where $S\left( \hat{\rho}\right) =-\mathrm{tr}\left( \hat{\rho}\log \hat{\rho
}\right) $ is von Neumann entropy of $\hat{\rho}$. One can see that $S\left(
\rho _{A}\right) =S\left( \rho _{B}\right) $. Although the entanglement
measure of a pure state is always zero, its reduced statistical operators
have nonzero entanglement measures.

For a the pure state $\left\vert \Psi \right\rangle
_{AB}=\sum_{s=1}^{4}\upsilon _{s}\left\vert \Theta \right\rangle _{s}$, we
obtain:
\begin{eqnarray*}
	&&\hat{\rho}_{AB}=\ _{AB}|\Psi \rangle \langle \Psi |\ _{AB}=\left[ \upsilon
	_{1}\left\vert 00\right\rangle +\upsilon _{2}\left\vert 01\right\rangle
	+\upsilon _{3}\left\vert 10\right\rangle +\upsilon _{4}\left\vert
	11\right\rangle \right] \\
	&&\times \left[ \upsilon _{1}^{\ast }\langle 00|+\upsilon _{2}^{\ast
	}\langle 01|+\upsilon _{3}^{\ast }\langle 10|+\upsilon _{4}^{\ast }\langle
	11|\right] \ .
\end{eqnarray*}
Then
\begin{align}
	& \ \hat{\rho}_{A}=\ _{B}\langle 0|\hat{\rho}_{AB}\left\vert 0\right\rangle
	_{B}+\ _{B}\langle 1|\hat{\rho}_{AB}\left\vert 1\right\rangle _{B}=|\upsilon
	_{1}|^{2}\left\vert 0\right\rangle \langle 0|+\upsilon _{1}\upsilon
	_{3}^{\ast }\left\vert 0\right\rangle \langle 1|  \notag \\
	& +|\upsilon _{2}|^{2}\left\vert 0\right\rangle \langle 0|+\upsilon
	_{2}\upsilon _{4}^{\ast }\left\vert 0\right\rangle \langle 1|+\upsilon
	_{3}\upsilon _{1}^{\ast }\left\vert 1\right\rangle \langle 0|+|\upsilon
	_{3}|^{2}\left\vert 1\right\rangle \langle 1|+\upsilon _{4}\upsilon
	_{2}^{\ast }\left\vert 1\right\rangle \langle 0|  \notag \\
	& +|\upsilon _{4}|^{2}\left\vert 1\right\rangle \langle 1|,\ \rho
	_{11}^{\left( A\right) }=|\upsilon _{1}|^{2}+|\upsilon _{2}|^{2},\ \rho
	_{12}^{\left( A\right) }=\upsilon _{1}\upsilon _{3}^{\ast }+\upsilon
	_{2}\upsilon _{4}^{\ast }\ ,  \notag \\
	& \ \rho _{21}^{\left( A\right) }=\upsilon _{3}\upsilon _{1}^{\ast
	}+\upsilon _{4}\upsilon _{2}^{\ast },\ \ \rho _{22}^{\left( A\right)
	}=|\upsilon _{3}|^{2}+|\upsilon _{4}|^{2}\ .  
	\label{21}
\end{align}
Calculating the entanglement measure, we may use eigenvalues of the matrix 
$\hat{\rho}_{A}$,
\begin{align}
	& \hat{\rho}^{\left( A\right) }P_{a}=\lambda _{a}P_{a},\ \lambda _{a}=\frac{1
	}{2}\left( |\upsilon _{1}|^{2}+|\upsilon _{2}|^{2}+|\upsilon
	_{3}|^{2}+|\upsilon _{4}|^{2}+\left( -1\right) ^{a}y\right) \ ,  \notag \\
	& P_{a}=\left( 
	\begin{array}{c}
		\frac{|\upsilon _{1}|^{2}+|\upsilon _{2}|^{2}-|\upsilon _{4}|^{2}-|\upsilon
			_{3}|^{2}+\left( -1\right) ^{a}y}{2\left( \upsilon _{3}\upsilon _{1}^{\ast
			}+\upsilon _{4}\upsilon _{2}^{\ast }\right) } \\ 
		1
	\end{array}
	\right) ,\ M\left( \left\vert \Psi \right\rangle _{AB}\right)
	=-\sum_{a=1,2}\lambda _{a}\log _{2}\lambda _{a}=H(z)\ ,  \notag \\
	& H(z)=-\left[ z\log _{2}z+\left( 1-z\right) \log _{2}(1-z)\right] ,\ \
	z=\left( 1+y\right) /2\ ,  \notag \\
	\ y& =\sqrt{\left( |\upsilon _{1}|^{2}+|\upsilon _{2}|^{2}-|\upsilon
		_{4}|^{2}-|\upsilon _{3}|^{2}\right) ^{2}+4\left\vert \upsilon _{1}\upsilon
		_{3}^{\ast }+\upsilon _{2}\upsilon _{4}^{\ast }\right\vert ^{2}},\ \left(
	0\log _{2}0\equiv 0\right) \ .  
	\label{20}
\end{align}
Function $H\left( x\right) $ is the so-called binary entropy function; see
e.g. Ref. \cite{Bennet2,Woott}.

\subsection{Entanglement of photons with anti-parallel polarizations in the
	model under consideration\label{SS2.2}}

Here using solutions of the model under consideration constructed in Sec. 
\ref{S1}, we study the entanglement of photons with anti-parallel
polarizations by the electron medium and by the external magnetic field.

Consider the state $\left\vert \Phi _{\mathrm{q-ph}}\right\rangle$ 
describing quasi-photons, with different frequencies and with anti-parallel
polarizations, $\lambda _{1}$ and $\lambda _{2}\neq \lambda _{1}$, i.e., 
$N_{1,3-\lambda _{1}}=N_{2,3-\lambda _{2}}=0,$ $N_{1,\lambda
_{1}}=N_{2,\lambda _{2}}=1$:
\begin{equation}
	\left\vert \Phi _{\mathrm{q-ph}}(\lambda _{1},\lambda _{2})\right\rangle =
	\hat{c}_{1,\lambda _{1}}^{\dagger }\hat{c}_{2,\lambda _{2}}^{\dagger
	}\left\vert 0\right\rangle _{\mathrm{c}},\ \ \left\vert
	0\right\rangle _{\mathrm{c}}=\left\vert 0\right\rangle _{\mathrm{c}
		_{1}}\otimes \left\vert 0\right\rangle _{\mathrm{c}_{2}},\ \ \hat{c}
	_{s,\lambda }\left\vert 0\right\rangle _{\mathrm{c}}=0,\ \ s=1,2\ .
	\label{f.0}
\end{equation}

With account taken of Eqs. (\ref{ac1}) and (\ref{16c}) we can see that the
last equation (\ref{f.0}) implies for small $\varepsilon $:
\begin{equation}
	\left\vert 0\right\rangle _{\mathrm{c}}=\left\vert 0\right\rangle -
	\frac{\omega _{0}\sqrt{\varepsilon }}{2eB}\left[ \sum_{s=1,2}\frac{i\hat{a}
		_{s,1}^{\dagger }+\hat{a}_{s,2}^{\dagger }}{\sqrt{\kappa _{s}}\left( \omega
		_{0}+\kappa _{s}\right) }\right] \hat{a}_{0}^{\dagger }\left\vert
	0\right\rangle +O(\varepsilon )\ .  
	\label{v.1}
\end{equation}
Then, it follows from Eq. (\ref{v.1}) that $\left\vert 0\right\rangle _{
	\mathrm{c}}=\left\vert 0\right\rangle +O(\sqrt{\varepsilon })$.
Taking into account expansion (\ref{ac1}) for state (\ref{f.0}), we obtain:
\begin{eqnarray}
	&&\left\vert \Phi _{\mathrm{q-ph}}(\lambda _{1},\lambda _{2})\right\rangle =
	\hat{c}_{1,\lambda _{1}}^{\dagger }\hat{c}_{2,\lambda _{2}}^{\dagger
	}\left\vert 0\right\rangle +O(\sqrt{\varepsilon })=\sum_{s,s^{\prime
		}=1,2}\sum_{\lambda ,\lambda ^{\prime }=1,2}u_{s,\lambda ;2,\lambda
		_{2}}u_{s,\lambda ;1,\lambda _{1}}\hat{a}_{s^{\prime },\lambda ^{\prime
	}}^{\dagger }\hat{a}_{s,\lambda }^{\dagger }\left\vert 0\right\rangle  \notag
	\\
	&&+u_{0,1;2,\lambda _{2}}\sum_{s=1,2}\sum_{\lambda =1,2}u_{s,\lambda
		;1,\lambda _{1}}\hat{a}_{s,\lambda }^{\dagger }\hat{a}_{0}^{\dagger
	}\left\vert 0\right\rangle +\left( \sum_{s=0,1,2}\sum_{\lambda
		=1,2}u_{s,\lambda ;2,\lambda _{2}}v_{s,\lambda ,1,\lambda _{1}}^{\ast
	}\right) \left\vert 0\right\rangle \ +O(\sqrt{\varepsilon })\ ,  \notag \\
	&&\left\Vert \Phi _{\mathrm{q-ph}}(\lambda _{1},\lambda _{2})\right\Vert
	^{2}=1+O(\sqrt{\varepsilon })\ .  
	\label{f.2}
\end{eqnarray}

We believe that the corresponding free photon nonentangled beam after
passing through the macro region, which consists of the electron media in
the presence of the magnetic field, is deformed to this form, and there
exists an analyzer detecting a two photon state for measuring the
entanglement of the initial free photons. The two photon state 
$|\tilde{\Phi}_{\mathrm{q-ph}}(\lambda _{1},\lambda _{2})\rangle$ is
represented by the first term in Eq. (\ref{f.2}),
\begin{equation}
	\left\vert \tilde{\Phi}_{\mathrm{q-ph}}(\lambda _{1},\lambda
	_{2})\right\rangle =D\sum_{s,s^{\prime }=1,2}\sum_{\lambda ,\lambda ^{\prime
		}=1,2}u_{s,\lambda ;2,\lambda _{2}}u_{s^{\prime },\lambda ^{\prime
		};1,\lambda _{1}}\hat{a}_{s^{\prime },\lambda ^{\prime }}^{\dagger }\hat{a}
	_{s,\lambda }^{\dagger }\left\vert 0\right\rangle ,\ \ \left\Vert \tilde{\Phi
	}_{\mathrm{q-ph}}\right\Vert ^{2}=1\ ,  
	\label{f3}
\end{equation}
where $D$ is a normalization factor. It follows from Eq. (\ref{16c}) that 
$u_{s,\lambda ;2,\lambda _{2}}u_{s,\lambda ^{\prime };1,\lambda _{1}}=O(
\varepsilon \left( \Delta \kappa \right) ^{-1})$. Then at $\Delta
\kappa =\left\vert \kappa _{2}-\kappa _{1}\right\vert \gg 1$, the two photon
state (\ref{f3}) can be reduced to the following form:
\begin{eqnarray}
	&&\left\vert \tilde{\Phi}_{\mathrm{q-ph}}(\lambda _{1},\lambda
	_{2})\right\rangle =D\sum_{\lambda ,\lambda ^{\prime }}\left[ u_{1,\lambda }
	\tilde{u}_{2,\lambda ^{\prime }}+u_{2,\lambda ^{\prime }}\tilde{u}
	_{1,\lambda }\right] \hat{a}_{1,\lambda }^{+}\hat{a}_{2,\lambda ^{\prime
	}}^{+}\left\vert 0\right\rangle +O\left( \left( \Delta \kappa \right)
	^{-1}\right) \ ,  \notag \\
	&&u_{s,\lambda }=u_{s,\lambda ;1,\lambda _{2}},\ \ \tilde{u}_{s,\lambda
	}=u_{s,\lambda ;2,\lambda _{1}}\ .  
	\label{e2}
\end{eqnarray}
In terms of the computational basis,
\begin{equation*}
	\left\vert \vartheta _{1}\right\rangle =a_{1,1}^{+}a_{2,1}^{+}\left\vert
	0\right\rangle ,\ \left\vert \vartheta _{2}\right\rangle
	=a_{1,1}^{+}a_{2,2}^{+}\left\vert 0\right\rangle ,\ \left\vert \vartheta
	_{3}\right\rangle =a_{1,2}^{+}a_{2,1}^{+}\left\vert 0\right\rangle ,\
	\left\vert \vartheta _{4}\right\rangle =a_{1,1}^{+}a_{2,1}^{+}\left\vert
	0\right\rangle \ ,
\end{equation*}
the state (\ref{e2}) can be rewritten as follows:
\begin{eqnarray}
	&&\left\vert \tilde{\Phi}_{\mathrm{q-ph}}(\lambda _{1},\lambda
	_{2})\right\rangle =D\sum_{j=1}^{4}\upsilon _{j}\left\vert \vartheta
	_{j}\right\rangle ,\ \ D=\left( \sum_{i=1}^{4}|\upsilon _{i}|^{2}\right)
	^{-1/2}\ ,  \notag \\
	&&\upsilon _{1}=u_{1,1}\tilde{u}_{2,1}+u_{2,1}\tilde{u}_{1,1},\ \ \upsilon
	_{4}=\upsilon _{1}\ ,  \notag \\
	&&\upsilon _{2}=u_{1,1}\tilde{u}_{2,2}+u_{2,2}\tilde{u}_{1,1},\ \ \upsilon
	_{3}=-\upsilon _{2}\ ,  
	\label{e4}
\end{eqnarray}

Let us calculate the entanglement measure $M(\lambda _{1},\lambda
_{2})=M(| \tilde{\Phi}_{\mathrm{q-ph}}(\lambda _{1},\lambda
_{2})\rangle)$ of the state $|\tilde{\Phi}_{\mathrm{q-ph}
}(\lambda _{1},\lambda _{2})\rangle$ as the von Neumann entropy of
the reduced density operator $\hat{\rho}^{\left( 1\right) }$ of the
subsystem of the first photon,
\begin{eqnarray}
	&&\ M(\lambda _{1},\lambda _{2})=-\mathrm{tr}\left( \hat{\rho}^{\left(
		1\right) }\log _{2}\hat{\rho}^{\left( 1\right) }\right) =-\sum_{a=1,2}\mu
	_{a}\log _{2}\mu _{a}\   \notag \\
	&&\ =-\left[ z\log _{2}z+\left( 1-z\right) \log _{2}(1-z)\right] ,\ z=\left(
	1+y\right) /2\ ,  
	\label{e.5}
\end{eqnarray}
where $\mu _{a}$, $a=1,2$, are eigenvalues of the operator $\hat{\rho}
^{\left( 1\right) }$, and
\begin{equation*}
	y=\sqrt{\left( |\upsilon _{1}|^{2}+|\upsilon _{2}|^{2}-|\upsilon
		_{4}|^{2}-|\upsilon _{3}|^{2}\right) ^{2}+4\left\vert \upsilon _{1}\upsilon
		_{3}^{\ast }+\upsilon _{2}\upsilon _{4}^{\ast }\right\vert ^{2}}.
\end{equation*}
In fact, we have to calculate the quantity $y$ to obtain the entanglement
measure $M(\lambda _{1},\lambda _{2})$. At small $\varepsilon $ they read:
\begin{eqnarray}
	&&y=1-\beta \varepsilon ^{4}+O(\varepsilon ^{5}),\ \ \beta =\frac{\left( 
		\bar{p}_{0}/\Delta \kappa \right) ^{4}}{8(\omega _{0}+(-1)^{\lambda
			_{1}}\kappa _{1})^{2}(\omega _{0}+(-1)^{\lambda _{2}}\kappa _{2})^{2}}\,,
	\label{e5} \\
	&&M(\lambda _{1},\lambda _{2})=-2\beta \varepsilon ^{4}\log _{2}\varepsilon +
	\frac{\beta }{2\ln 2}\left( 1-\ln \frac{\beta }{2}\right) \varepsilon
	^{4}+O(\varepsilon ^{5})\ .  \notag
\end{eqnarray}

Further, it is convenient for us to choose a reference frame relative to
which the momentum $p_{3}$ of electrons in the charged medium is zero, $
p_{3}=0$. Then the quantity $\omega _{0}$ is related to the magnetic field 
$B$ as:
\begin{equation}
	\omega _{0}=\frac{eB}{\sqrt{2eB\left( N_{0}+1/2\right) +m^{2}}}\ .
	\label{e6}
\end{equation}

We note that the quantity $y$ given by Eq. (\ref{e5}) is singular, if
\begin{equation}
	\omega _{0}=\left\{ 
	\begin{array}{l}
		\kappa _{1},\ \ \mathrm{if}\ \ \lambda _{1}=1,\ \ \lambda _{2}=2\  \\ 
		\kappa _{2},\ \ \mathrm{if}\ \ \lambda _{1}=2,\ \ \lambda _{2}=1
	\end{array}
	\right. \ .  
	\label{e6b}
\end{equation}
The corresponding to such $\omega _{0}$ strengths of the magnetic field $B$,
will be called resonant ones. There exist two such resonant values, $B=B_{1}$
at $\omega _{0}=\kappa _{1}$ for $\lambda _{1}=1$ and $B=B_{2}$ at $\omega
_{0}=\kappa _{2}$ for $\lambda _{1}=2$:
\begin{eqnarray}
	B_{1} &=&\frac{\kappa _{1}}{e}\sqrt{(N_{0}+1/2)^{2}\kappa _{1}^{2}+m^{2}}
	+\left( N_{0}+1/2\right) \kappa _{1}^{2}\ ,  \notag \\
	B_{2} &=&\frac{\kappa _{2}}{e}\sqrt{(N_{0}+1/2)^{2}\kappa _{2}^{2}+m^{2}}
	+\left( N_{0}+1/2\right) \kappa _{2}^{2}\ .  \label{e7}
\end{eqnarray}

When $B=B_{1}$, the expansions
\begin{eqnarray}
	\tau _{1,2} &=&\kappa _{1}+\frac{\kappa _{1}\varepsilon }{2(\kappa
		_{1}+\kappa _{2})eB_{1}}+O(\varepsilon ^{2})\ ,  \notag \\
	\tau _{2,1} &=&\kappa _{2}-\sqrt{\frac{\kappa _{1}\varepsilon }{2eB_{1}}}
	+O(\varepsilon )  
	\label{e8}
\end{eqnarray}
take place. Similarly, when $B=B_{2}$, the expansions
\begin{eqnarray}
	\tau _{1,2} &=&\kappa _{1}+\frac{\kappa _{2}\varepsilon }{2(\kappa
		_{1}+\kappa _{2})eB_{2}}+O(\varepsilon ^{2})\ ,  \notag \\
	\tau _{2,1} &=&\kappa _{2}-\sqrt{\frac{\kappa _{2}\varepsilon }{2eB_{2}}}
	+O(\varepsilon )  
	\label{e8b}
\end{eqnarray}
hold. They have a different character than the one given by Eqs. (\ref{15})
for similar roots. We suppose that at $B=B_{1}$ or $B=B_{2}$ analytical
properties of roots (\ref{15}) change as functions of the parameter 
$\varepsilon$.

From (\ref{e8}) and (\ref{e8b}), we find that when a resonant value $B$ is
reached, the entanglement manifests itself already in a lower order in 
$\varepsilon$ compared to expression (\ref{e5}). For $B=B_{1}$, we have: 
\begin{eqnarray}
	&&y=1-\delta _{1}\varepsilon ^{3}+O(\varepsilon ^{4}),\ \delta _{1}=\frac{1}{
		4(\Delta \kappa )^{4}(\kappa _{1}+\kappa _{2})^{2}}\left( \frac{\kappa _{1}}{
		eB_{1}}\right) ^{3}\ ,  \notag \\
	&&M(\lambda _{1},\lambda _{2})=-\frac{3}{2}\delta _{1}\varepsilon ^{3}\log
	_{2}\varepsilon +\frac{\delta _{1}}{2\ln 2}\left( 1-\ln \frac{\delta _{1}}{2}
	\right) \varepsilon ^{3}+O(\varepsilon ^{4})\ ,  
	\label{e9}
\end{eqnarray}
whereas for $B=B_{2},$ we obtain:
\begin{eqnarray}
	&&y=1-\delta _{2}\varepsilon ^{3}+O(\varepsilon ^{4}),\ \ \delta _{2}=\frac{1
	}{4(\Delta \kappa )^{4}(\kappa _{1}+\kappa _{2})^{2}}\left( \frac{\kappa _{2}
	}{eB_{2}}\right) ^{3}\ ,  \notag \\
	&&M(\lambda _{1},\lambda _{2})=-\frac{3}{2}\delta _{2}\varepsilon ^{3}\log
	_{2}\varepsilon +\frac{\delta _{2}}{2\ln 2}\left( 1-\ln \frac{\delta _{2}}{2}
	\right) \varepsilon ^{3}+O(\varepsilon ^{4})\ .  
	\label{e10}
\end{eqnarray}

It can be seen that if the photon polarizations are the same, 
$\lambda_{1}=\lambda _{2}$ then the entanglement measure is equal to zero, 
$M(1,1)=M(2,2)=0$.

\section{Illustrative numerical calculations and some final remarks\label{S3}}

In our numerical calculations, we consider all the electrons in the charge
medium located on zero Landau level $N_{0}=0$ and the beam of two photons
with polarization $\lambda _{1}=2$ and $\lambda _{2}=1$. It follows from Eq.
(\ref{e5}) that the resonant entanglement is related to the frequency of
that photon whose polarization vector is directed along the $Ox$ axis at 
$B>0$. If you change the direction of the magnetic field, $B<0$, then the
resonant entanglement will be related to the frequency of that photon whose
polarization vector is directed along the $Oy$ axis. Therefore in the case
under consideration we have the resonant value of the magnetic field is 
$B=B_{2}$, see Eq. (\ref{e7}).

On the first plot the entanglement measure $M$ $(2,1)$ is calculated as a
function of the magnetic field $B$ for the fixed first photon frequency 
$\nu_{1}=10^{3}\ \mathrm{nm}$, and different second photon frequency 
$\nu_{2}=2\pi \kappa _{2}^{-1}$. The electron density is chosen to be 
$\rho=10^{14}\mathrm{el\ m}^{-3}$.

\begin{figure}[h]
	\centering
	\includegraphics[width=0.8\textwidth]{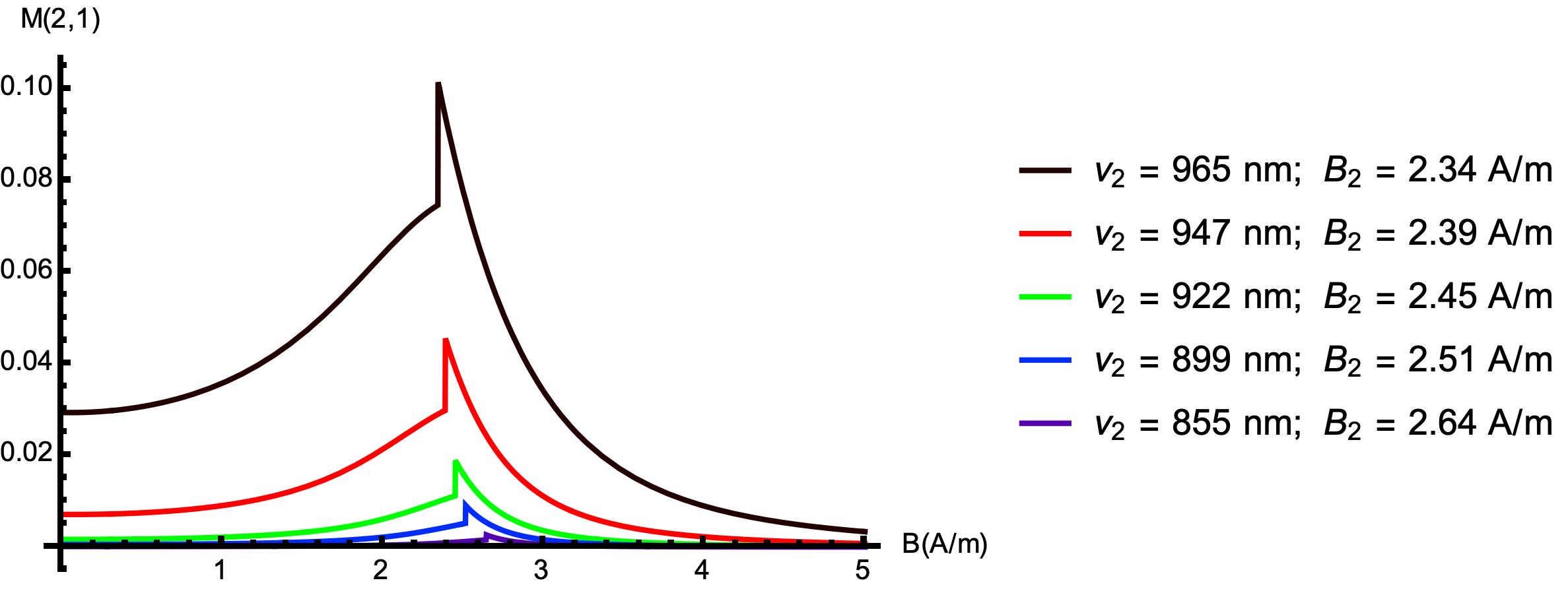}
	\caption{The entanglement measure as a function of the magnetic field.}
	\label{graph1}
\end{figure}

We see that the entanglement measure increases with increasing the magnetic
field strength $B<B_{\mathrm{2}}$. When the magnetic field reaches its
resonant value $B=B_{2}$, the entanglement measure experiences a jump. A
further increase in the magnetic field $B>B_{\mathrm{2}}$ leads to a smooth
decrease in the entanglement. We also see that the entanglement measure
decreases as the difference in photon frequencies increases. In the work 
\cite{360} the entanglement of two photons in the absence of a magnetic
field was considered and it was shown that the measure of the entanglement
is the same for $\lambda _{1}=1$, $\lambda _{2}=2$ and $\lambda _{1}=2$, 
$\lambda _{2}=1$. Here it is demonstrated that the presence of the magnetic
field removes the degeneracy in photon polarizations and the entanglement
measure depends on the direction of photon polarizations in the beam.
Increasing the magnetic field strength increases entanglement, as long as
the magnetic field value is below a certain resonant value, which is
determined by the frequency of the photon having polarization $\lambda =1$,
see Eq. (\ref{e7}). The resonant value of $B$ increases with decreasing of
photon frequencies. But these values are not large, for example, for photons
with frequencies $\nu _{2}$ corresponding to the ultraviolet range $380$ 
\textrm{nm} -- $10$ \textrm{nm}, the resonant values range from 
$6$ \textrm{A/m} to $225$ \textrm{A/m}.

On the second plot the entanglement measure $M$ $(2,1)$ is calculated as a
function of the electron medium density for the fixed first photon frequency 
$\nu _{1}=10^{3}\ \mathrm{nm,}$ and different second photon frequencies 
$\nu_{2}$. The magnetic field $B$ is chosen to be $B=2$ \textrm{A/m} which is
less the corresponding resonant values.

\begin{figure}[h]
	\centering
	\includegraphics[width=0.8\textwidth]{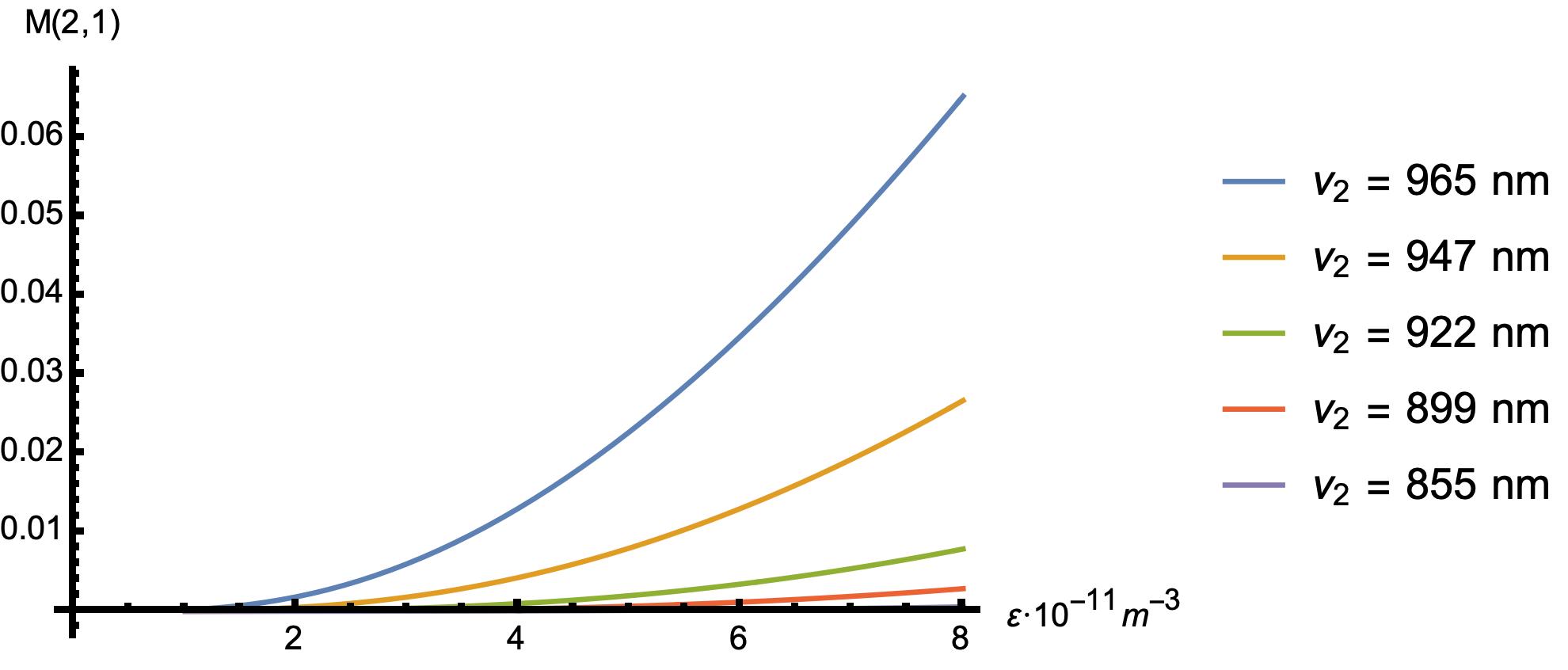}
	\caption{The entanglement measure $M(2,1)$ as a function of the
		electron medium density.}
	\label{graph2}
\end{figure}

Note that the measure of entanglement increases with increasing
density of the electronic medium and with increasing of pre-resonance values
of the magnetic field. In our calculations the entanglement measure does not
exceed $0.1$. However, such a magnitude of the entanglement is usual in
laboratory experiments, for example, similar magnitudes appear when an
entangled biphoton Fock state of photons is scattered inside an optical
cavity; see Refs. \cite{Li2019, Malatesta2023}.

We stress that performed numerical calculations are intended to illustrate
the existence of a possible resonant entanglement within the framework of
the chosen model and the approximations made. On the other hand, if our
consideration motivates possible experiments to detect the effect of
resonant entanglement then there may be an incentive to refine the
corresponding model, in particular, the analytical formulas (\ref{e5}), 
(\ref{e9}) and (\ref{e10}) under weaker restrictions on the density of the
electron medium and frequencies of the photons.

\section{Acknowledgments}

The work is supported by Russian Science Foundation, grant No. 19-12-00042. D.M.G. thanks CNPq for permanent support.

\section{Appendix. Diagonalization of the operator $\hat{H}_{\mathrm{\protect\chi}}\left( 0\right) $}\label{A}

Here we are going to construct a linear canonical transformation from the
operators $a_{s,\lambda }^{\dagger }$ and $\hat{a}_{s,\lambda }$, $s=0,1,2$,
to some new Bose operators $\hat{c}_{s,\lambda }^{\dagger }$ and $\hat{c}
_{s,\lambda }$, $s=0,1,2$,
\begin{eqnarray}
	&&\hat{c}_{s,\lambda }=\sum_{s=0,1,2;}\sum_{\lambda =1,2}u_{s^{\prime
		},\lambda ^{\prime };s,\lambda }^{\ast }\hat{a}_{s^{\prime },\lambda
		^{\prime }}+v_{s^{\prime },\lambda ^{\prime };s,\lambda }\hat{a}_{s^{\prime
		},\lambda ^{\prime }}^{\dagger }\ ,\ \ \hat{c}_{0,\lambda }=\hat{c}
	_{0}\delta _{\lambda ,1}\ ,  \notag \\
	&&\hat{c}_{s,\lambda }^{\dagger }=\sum_{s=0,1,2;}\sum_{\lambda
		=1,2}u_{s^{\prime },\lambda ^{\prime };s,\lambda }\hat{a}_{s^{\prime
		},\lambda ^{\prime }}^{\dagger }+v_{s^{\prime },\lambda ^{\prime };s,\lambda
	}^{\ast }\hat{a}_{s^{\prime },\lambda ^{\prime }}\ ,\ \ \hat{c}_{0,\lambda
	}^{\dagger }=\hat{c}_{0}^{\dagger }\delta _{\lambda ,1}\ ,  \label{ac1} \\
	&&\left[ \hat{c}_{s,\lambda },\hat{c}_{s^{\prime },\lambda ^{\prime }}\right]
	=\left[ \hat{c}_{s,\lambda }^{\dagger },\hat{c}_{s^{\prime },\lambda
		^{\prime }}^{\dagger }\right] =0\ ,\ \ \left[ \hat{c}_{s,\lambda },\hat{c}
	_{s^{\prime },\lambda ^{\prime }}^{\dagger }\right] =\delta _{s,s^{\prime
	}}\delta _{\lambda ,\lambda ^{\prime }}\ ,  
	\label{comm1}
\end{eqnarray}
in terms of which operator (\ref{a2}) takes the following diagonal form:
\begin{equation}
	\hat{H}_{\mathrm{\chi }}(0)=\sum_{s=0,1,2}\sum_{\lambda =1,2}\tau
	_{s,\lambda }\hat{c}_{s,\lambda }^{\dagger }\hat{c}_{s,\lambda }+\tilde{H}
	_{0}\ ,\ \ \tau _{0,\lambda }=\tau _{0}\delta _{\lambda ,1}\ ,\ \ \tau
	_{s,\lambda }>0\ ,\ \ \tilde{H}_{0}=\mathrm{const}\ .  
	\label{a0}
\end{equation}

It is known that for the linear transformation (\ref{ac1}) to be canonical
(namely, Eqs. (\ref{comm1}) hold true) is reduced to the matrices $u=\left(
u_{s,\lambda ;s^{\prime },\lambda ^{\prime }}\right) $ and $v=\left(
v_{s,\lambda ;s^{\prime },\lambda ^{\prime }}\right) $ must satisfy the set
of equations
\begin{equation}
	uu^{\dagger }-vv^{\dagger }=1,\ \ vu^{T}=uv^{T}\ ;  
	\label{uv1}
\end{equation}
(see Ref. \cite{berezin}). We note that we are looking for the those
canonical transformations that diagonalizes the Hamiltonian $\hat{H}_{
\mathrm{\chi }}(0)$ transforming it to form (\ref{a0}). In this case with
account taken of Eqs. (\ref{comm1}) and (\ref{a0}) we obtain:
\begin{equation}
	\left[ \hat{H}_{\mathrm{\chi }}(0),\hat{c}_{s,\lambda }\right] =-\tau
	_{s,\lambda }\hat{c}_{s,\lambda }\ .  
	\label{comm2}
\end{equation}
Substituting Eqs. (\ref{a2}) and (\ref{ac1}) into Eq. (\ref{comm2}), we
obtain the following system of equations
\begin{eqnarray}
	&&\left( A-\tau \right) u^{+}-Bv^{T}=0,\ \ \left( A+\tau \right)
	v^{T}-B^{+}u^{+}=0\ ,  \notag \\
	&&A=\left( A_{s,\lambda ;s^{\prime },\lambda ^{\prime }}\right) ,\ \
	B=\left( B_{s,\lambda ;s^{\prime },\lambda ^{\prime }}\right) ,\ \ \tau
	=\left( \tau _{s,\lambda }\right) \ .  
	\label{a4b}
\end{eqnarray}

In contrast to Eqs. (\ref{uv1}), it is linear in the matrices $u$ and $v$
which allows one relatively easy its analysis. One can see that system 
(\ref{a4b}) is joint if positive numbers $\tau =(\tau _{s,\lambda })$ for each
possible set $s=0,1,2$ and $\lambda =1,2$ satisfy the equations:
\begin{eqnarray}
	&&\left( \sum_{s=1,2}\frac{\epsilon }{\tau ^{2}-\kappa _{s}^{2}}-1+\frac{
		\omega }{\tau }\right) \left( \sum_{s^{\prime }=1,2}\frac{\epsilon }{\tau
		^{2}-\kappa _{s^{\prime }}^{2}}-1-\frac{\omega }{\tau }\right) =0,\ \
	\epsilon \neq 0\ ,  \label{a6} \\
	&&\left( \tau -\kappa _{1}\right) ^{2}(\tau -\kappa _{2})^{2}\left[ \left( 
	\frac{\omega }{\tau }\right) ^{2}-1\right] =0,\ \ \epsilon =0\ .  
	\label{a6b}
\end{eqnarray}
We now suppose that roots $\tau$ of Eq. (\ref{a6}) are at the same time,
solutions of the equation
\begin{equation}
	\sum_{s^{\prime }=1,2}\frac{\epsilon }{\tau _{s,\lambda }^{2}-\kappa
		_{s^{\prime }}^{2}}=1+\frac{(-1)^{\lambda }\omega }{\tau _{s,\lambda }}
	\label{a8}
\end{equation}
with the initial conditions:
\begin{equation}
	\left. \tau _{0,1}\right\vert _{\epsilon \rightarrow 0}=\omega ,\ \ \left.
	\tau _{s,1}\right\vert _{\epsilon \rightarrow 0}=\kappa _{s},\ \ \left. \tau
	_{s,2}\right\vert _{\epsilon \rightarrow 0}=\kappa _{s}\ .  
	\label{a8b}
\end{equation}
Thus, we define three positive roots $\tau _{0,1}$, $\tau_{1,1}$, and $\tau
_{2,1}$ for $\lambda =1$ and two different positive roots $\tau _{1,1}$ and 
$\tau _{1,2}$ for $\lambda =2$. Due to condition (\ref{a8b}), these five
roots must be reduced to roots of Eq. (\ref{a6b}) as $\epsilon \rightarrow 0$.

Solving now Eqs. (\ref{a4b}), we obtain the set:
\begin{eqnarray}
	&&u_{s,\lambda ;k,\sigma }=\left[ \left( \sqrt{\frac{\tau _{k,\sigma }}{
			\kappa _{s}}}+\sqrt{\frac{\kappa _{s}}{\tau _{k,\sigma }}}\right) \frac{
		\delta _{\lambda ,1}-i\left( -1\right) ^{\sigma }\delta _{\lambda ,2}}{
		2\left( \tau _{k,\sigma }^{2}-\kappa _{s}^{2}\right) }\left( 1-\delta
	_{s,0}\right) -\delta _{s,0}\delta _{1,\lambda }\delta _{1,\sigma }\sqrt{
		\frac{\omega }{\epsilon r_{k,\sigma }^{3}}}\right]\nonumber\\
	&&\qquad\qquad\times \left( 1-\delta _{\sigma
		,2}\delta _{k,0}\right) q_{k,\sigma }\ ,  \notag \\
	&&v_{s,\lambda ;k,\sigma }=\left[ \left( \sqrt{\frac{\tau _{k,\sigma }}{
			\kappa _{s}}}-\sqrt{\frac{\kappa _{s}}{\tau _{k,\sigma }}}\right) \frac{
		\delta _{\lambda ,1}+i\left( -1\right) ^{\sigma }\delta _{\lambda ,2}}{
		2\left( \tau _{k,\sigma }^{2}-\kappa _{s}^{2}\right) }\left( 1-\delta
	_{s,0}\right) -\delta _{s,0}\delta _{1,\lambda }\delta _{1,\sigma }\sqrt{
		\frac{\omega }{\epsilon r_{k,\sigma }^{3}}}\right]\nonumber\\
	&&\qquad\qquad\times\left( 1-\delta _{\sigma
		,2}\delta _{k,0}\right) q_{k,\sigma }\ .  
	\label{a9}
\end{eqnarray}
Substituting it into Eqs. (\ref{uv1}) and taking into account Eq. (\ref{a8}
), we derive the following expressions for the quantities $q_{k,\sigma }$:
\begin{equation*}
	q_{k,\sigma }=\left[ \frac{\left( -1\right) ^{\sigma -1}\omega }{\epsilon
		\tau _{k,\sigma }^{3}}+2\sum_{s=1,2}\left( \tau _{k,\sigma }^{2}-\kappa
	_{s}^{2}\right) ^{-2}\right] ^{-1/2}\ .
\end{equation*}
We can verify that the Eqs. $\det u\neq 0$ and $\det v\neq 0$ hold true such
that transformation (\ref{ac1}) is an reversible.

Let us return to Eq. (\ref{a0}) and finally determine the form of the
constant $\tilde{H}_{0}$. To this end, we consider the vacuum mean (with
respect to vacuum (\ref{a1a})) of the operator $\hat{H}_{\mathrm{\chi }}(0)$
in its initial form (\ref{a2}),
\begin{equation}
	\left\langle 0\left\vert \hat{H}_{\mathrm{\chi }}(0)\right\vert
	0\right\rangle =\frac{m^{2}}{2ng}+\frac{\omega }{2}+\frac{\epsilon }{2}
	\sum_{s=1,2}\kappa _{s}^{-1}\ .  
	\label{a14}
\end{equation}
With account taken of Eqs. (\ref{ac1}) and (\ref{a0}), we obtain:
\begin{equation}
	\left\langle 0\left\vert \hat{H}_{\mathrm{\chi }}(0)\right\vert
	0\right\rangle =\tilde{H}_{0}+\sum_{s,k=0,1,2}\sum_{\lambda ,\lambda
		^{\prime }=1,2}\tau _{k,\lambda ^{\prime }}\left\vert v_{s,\lambda
		;k,\lambda ^{\prime }}\right\vert ^{2}\ .  
	\label{a15}
\end{equation}
Comparing RHS of Eqs. (\ref{a14}) and (\ref{a15}), we obtain the constant 
$\tilde{H}_{0}$,
\begin{equation}
	\tilde{H}_{0}=\frac{m^{2}}{2ng}+\frac{\omega }{2}-\sum_{s,k=0,1,2}\sum_{
		\lambda ,\lambda ^{\prime }=1,2}\tau _{k,\lambda ^{\prime }}\left\vert
	v_{s,\lambda ;k,\lambda ^{\prime }}\right\vert ^{2}+\frac{\epsilon }{2}
	\sum_{s=1,2}\kappa _{s}^{-1}\ .  
	\label{a15b}
\end{equation}

\end{document}